\documentclass[aps,preprint,showpacs,preprintnumbers,amsmath,amssymb,a4paper,superscriptaddress]{revtex4}

\usepackage{graphicx}
\usepackage{dcolumn}
\usepackage{bm}

\begin{document}

\title{Effective Long-Range Interactions in Confined Curved Dimensions}
\author{P. Schmelcher}
\affiliation{Zentrum f\"{u}r Optische Quantentechnologien, Universit\"{a}t Hamburg, Luruper Chaussee 149, 22761 Hamburg, Germany}
\email{pschmelc@physnet.uni-hamburg.de}

\date{\today}

\begin{abstract}
We explore the effective long-range interaction of charged particles confined to a curved low-dimensional manifold using the example of
a helical geometry. Opposite to the Coulomb interaction in free space the confined particles experience a force which is oscillating
with the distance between the particles. This leads to stable equilibrium configurations and correspondingly induced bound
states whose number is tunable with the parameters of the helix. We demonstrate the existence of a plethora of equilibria of
few-body chains with different symmetry character that are allowed to freely move. An outline concerning the implications
on many-body helical chains is provided.
\end{abstract}


\maketitle

\section{Introduction}

Dimensionality and geometry often play a key role in physical systems and are responsible for many of their unusual properties. 
This holds in particular for condensed matter systems where celebrated effects such as the quantum hall effect \cite{qhe},
high temperature superconductivity \cite{htsc} or generally the physics of quantum phase transitions \cite{qpt} are 
connected or even intrinsically tight to the underlying dimensionality. In the case of cold and ultracold quantum gases \cite{pethick,pitaevskii},
the extensive control of the external as well as internal degrees of freedom of the atoms achieved during the past decades
allows to systematically tune the dimensionality of the confining traps thereby covering the complete crossover from zero-dimensional
dot-like ensembles via one-dimensional gases to two-dimensional layered structures. Furthermore, individual traps can be shaped
almost arbitrarily and arrays of traps can be prepared and manipulated by employing magnetic and/or optical fields 
\cite{weidemueller,schmiedmayer,oberthaler}. At the same time, the short-range interactions among ultracold atoms
can be adjusted covering weak to strongly correlated systems by using magnetic or optical Feshbach resonances \cite{chin,koehler,bauer}.
Long-range interactions of dipolar character, which can be realized by, e.g., atoms with a large magnetic dipole moment or
heteronuclear molecules exposed to electric fields, add a plethora of novel phenomena both on the mean-field and
strongly correlated many-body level \cite{lahaye}. Many of these effects, such as the roton-maxon spectrum and novel quantum phases
in optical lattices, are based on the anisotropic character of the dipolar interaction. In spite of these intriguing
findings, the underlying fundamental interactions (short-range, dipolar or even Coulomb interaction among e.g. ions) typically take on a very few specific 
appearances only. They are either overall repulsive or exhibit a single potential well. Asymptotically, i.e. for large distances, typically
a power law behaviour is encountered.

The question we pose here is whether dimensionality and geometry can cooperate in such a way that novel effective long-range
interactions emerge. To tackle this question, we consider in the present work long-range (Coulomb) interacting particles
that are confined in two out of three dimensions i.e. the particles are allowed
to move on a one-dimensional manifold (ODM) only. The latter possesses a nontrivial geometry in the sense of a non-vanishing 
curvature with respect to its embedding into three-dimensional space. Although the dynamics is one-dimensional, i.e. the two 
confined dimensions are not accessible dynamically, interactions take place via three-dimensional space. This introduces
effective one-dimensional interactions along the ODM seen by the particles which can exhibit peculiar properties, such as 
an oscillating short-range to long-range behaviour with an adjustable number of equilibrium configurations depending on the curvature and
near encounters of the ODM. Physically speaking, one might think of a two-dimensional tightly confining trap that freezes the dynamics of
two of the three degrees of freedom of the interacting particles and the remaining unconfined motional degree of freedom possesses a nonzero curvature. 
We perform a comprehensive analysis of the two-body effective interactions and dynamics and show the existence of 
few-body bound states for purely repulsive interactions in three dimensions. The latter leads to a plethora of stable many-body configurations of chains
in the ODM. In this context it is worthy to note that recently \cite{law} cold polar molecules confined to a helical optical lattice have been shown to exhibit an attractive
large-distance intermolecular interaction but the short-scale interaction is repulsive due to geometric constraints which leads to a zero-temperature
second-order liquid-gas transition at a critical applied electric field. 

\begin{figure}
\includegraphics[width=\columnwidth]{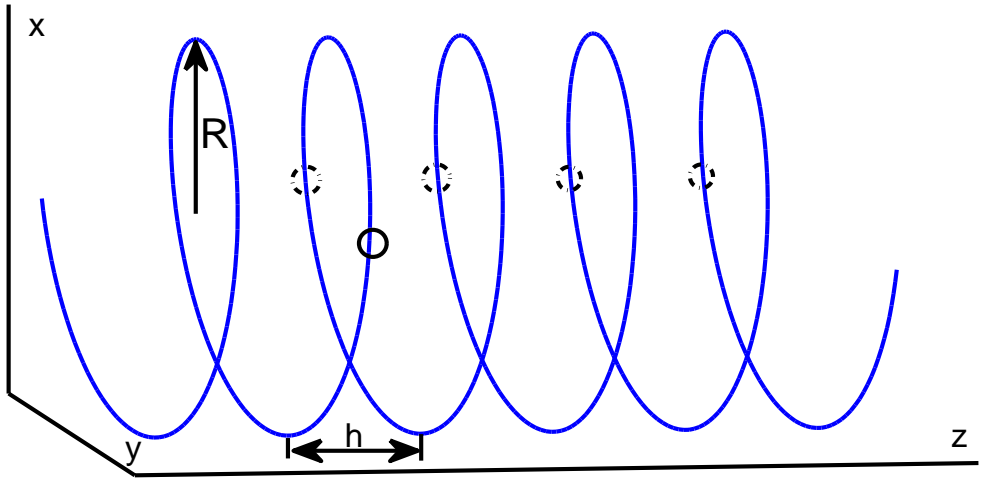}
\caption{\label{fig:fig1} Schematic illustration of the helix including the radius $R$ of the projection onto the
$x,y$-plane and the pitch $h$. Each pair of full and broken circles indicate the relative positions of two (equally charged) particles
for a corresponding stable equilibrium configuration.}
\end{figure}

\section{Lagrangian}

We consider charged particles interacting via the Coulomb potential that are confined to a one-dimensional manifold.
From the latter we require homogeneity and at the same time a nontrivial geometry in the sense of a non-vanishing curvature. One of the
basic setups fulfilling these criteria is a helix characterized by a constant curvature and a given pitch (see Fig.1). The charged particles are
allowed to move on the one-dimensional helix only but are interacting via three-dimensional space. Our starting-point is the 
Lagrangian ${\cal{L}} = {\cal{T}} - {\cal{V}}$ using three-dimensional Cartesian coordinates ${\bf{r}}_i$ and corresponding velocities ${\bf{\dot{r}}}_i$
where ${\cal{T}}=\sum_i \frac{m_i}{2} {\bf{\dot{r}}}_i^2$, ${\cal{V}}=\sum_{i > j} g_{ij}/|{\bf{r}}_i - {\bf{r}}_j|$ and $m_i$ are the kinetic and
potential energy and masses of the particles, respectively. $g_{ij}$ are the coupling constants of the interacting particles. 
Let us now specify for each particle $i$ its position on the helix
by introducing the corresponding (signed) path length $s_i$ along the helix
which is connected to the Cartesian coordinates via $x_i = R \sin \left( s_i/\alpha \right)$,
$y_i = R \cos \left( s_i/\alpha \right)$ and $z_i= \left( h/ 2\pi \right) \left(s_i/\alpha \right)$ where $R$ and $h$ are the 
radius of the circle, which is the orthogonal projection of the helix onto the $(x,y)$-plane, and the pitch, respectively.
Furthermore $\alpha = \sqrt{R^2 + \left(h/2\pi\right)^2}$ provides the proportionality factor between the path length and the azimuthal angle in the
$(x,y)$-plane i.e. $s_i = \alpha \cdot \phi_i$. The Lagrangian for the one-dimensional helical motion then takes on the appearance
\begin{eqnarray}
&{\cal{L}}& = \sum_{i=1}^{N} \frac{m_i}{2} {\dot{s}}_i^2 \\ \nonumber
&-& \sum_{i>j} \frac{g_{ij}}{\sqrt{2 R^2 \left( 1 - \cos \left( \frac{1}{\alpha} \left(s_i - s_j 
\right) \right) \right) + \left(\frac{h}{2 \pi \alpha} \right)^2 \left( s_i - s_j \right)^2 }}
\end{eqnarray}
From the above, it is evident that the kinetic energy has retained its original Cartesian form and the interaction potential 
depends only on $|s_i - s_j|$. Therefore, besides the energy, the total momentum is conserved
and it is possible to separate the center of mass and relative motion which is a consequence of the spatial homogeneity on the helix.
Introducing the center of mass $S=\left(\sum_{i=1}^{N} m_i s_i \right)/M$ and $(N-1)$ relative coordinates $\tilde{s}_i=s_{i+1}-s_i, i=1,...,N-1$
yields the following many-body Lagrangian
\begin{eqnarray}
&{\cal{L}}& = \frac{M}{2} {\dot{S}}^2 
+ \frac{m}{2} \sum_{i=1}^N \left(\frac{1}{N} \sum_{k=1}^{N-1} k {\dot{\tilde{s}}}_k - \sum_{l=i}^{N-1} {\dot{\tilde{s}}}_l \right)^2 \\ \nonumber
&-& \sum_{k=1}^{N-1} \sum_{n=k}^{N-1} \frac{g}{\sqrt{
\left(\frac{h}{2 \pi \alpha}\right)^2 \left( \sum_{i=k}^{n} {\tilde{s}}_i \right)^2
+ 2 R^2 \left( 1 - \cos \left( \frac{1}{\alpha} \sum_{i=k}^{n} {\tilde{s}}_i  \right) \right) }}
\label{manybodham}
\end{eqnarray}
where $M$ is the total mass ($g_{ij}=g$ and $g > 0$ as well as $m_i=m$). In the case of two particles this simplifies to
the integrable Lagrangian
\begin{eqnarray}
{\cal{L}}_2 = \frac{M}{2} {\dot{S}}^2 + \frac{m}{4} {{\dot{s}}}^2
- \frac{g}{\sqrt{\left(\frac{h}{2 \pi \alpha}\right)^2 s^2
+ 2 R^2 \left( 1 - \cos \left( \frac{s}{\alpha} \right) \right)}}
\label{twobodham}
\end{eqnarray}

where $s=s_2-s_1$ and we have omitted the tilde and will, from now on, disregard the uniform center of mass motion.
The last term of eq.(\ref{twobodham})
establishes an 'effective' interaction potential $V_{\mathrm{eff}}$ between the particles which are constrained to move on the helix. 
Note, that the variable $s$ encodes the distance of the two particles on the helix and therefore replaces the distance of
two particles in Cartesian space. The first term under the square root is reminescent of the radial dependence of the 
Coulomb potential. The second term, however,
has no counterpart in three-dimensional space and possesses an oscillatory behaviour. It therefore leads to a very 
unusual two-body interaction which will be analyzed in the following.

\section{Two-body problem}

In spite of the purely repulsive ($g>0$) Coulomb interaction in free space, $V_{\mathrm{eff}}$ can, depending on the parameters
$R,h$, not only be nonmonotonic with respect to its derivatives but even develop several local minima which qualify for a local oscillatory
dynamics. To explore this, let us analyze the necessary conditions for the occurence of extrema of $V_{\mathrm{eff}}$. 
A vanishing force $F_s = - \frac{dV_{\mathrm{eff}}}{ds} = 0$ leads to the implicit condition 
\begin{equation}
\frac{R^2}{\alpha} \sin\left(\frac{s}{\alpha}\right) + \left(\frac{h^2}{4 \pi^2 \alpha^2}\right) s = 0
\label{cond1}
\end{equation}
which can geometrically be interpreted as the intersection points of a sine-function and a straight line.
Eq.(\ref{cond1}) provides the $s$-values which correspond to the extrema of the relative positions of the two equally charged
particles on the helix for given parameters $R,h$. With varying parameters $R,h$ new extrema emerge if condition
(\ref{cond1}) is fulfilled and the sine-function is tangential to the straight line. This leads to explicit
conditions for the corresponding $s$-values and an implicit equation for the parameters

\begin{eqnarray}
s^2 = \alpha^2  \left( \left( \frac{2 \pi R}{h} \right)^4 -1 \right) \\
\cos \left( \sqrt{\left( \frac{2 \pi R}{h} \right)^4 -1} \right) = - \left( \frac{h}{2 \pi R} \right)^2
\end{eqnarray}

\begin{figure}
\includegraphics[width=0.8\columnwidth]{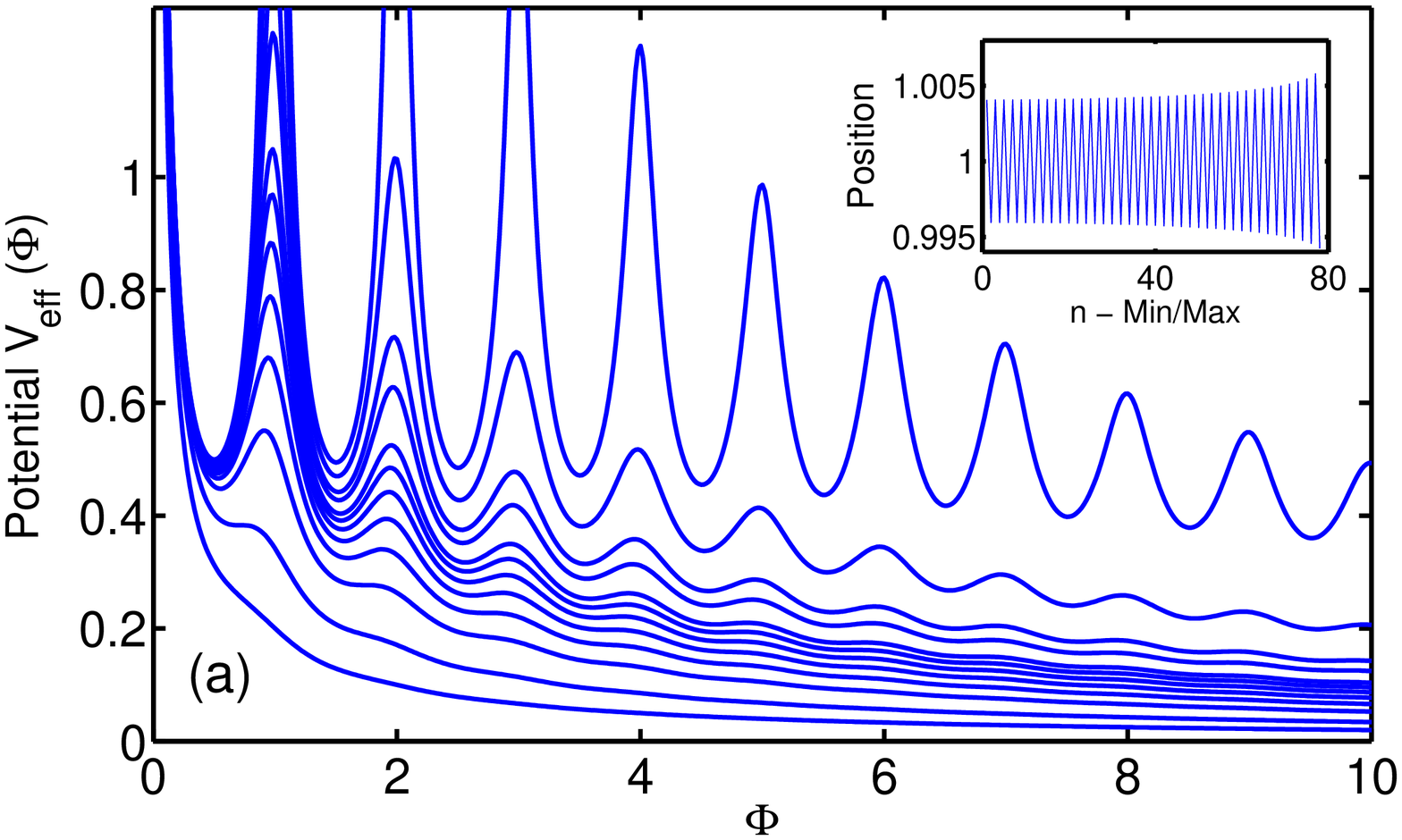}\\
\includegraphics[width=0.7\columnwidth]{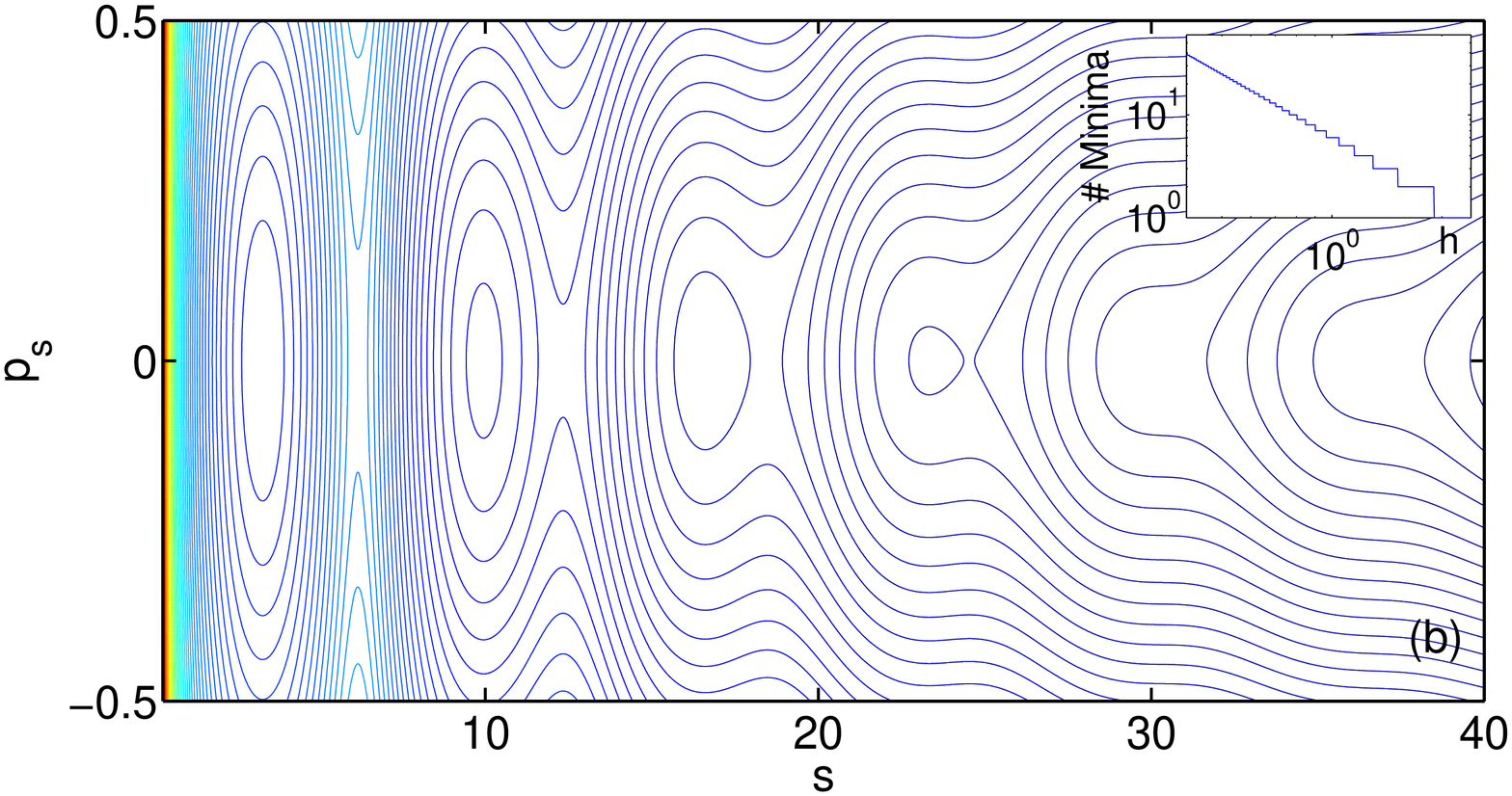}\\
\includegraphics[width=0.7\columnwidth]{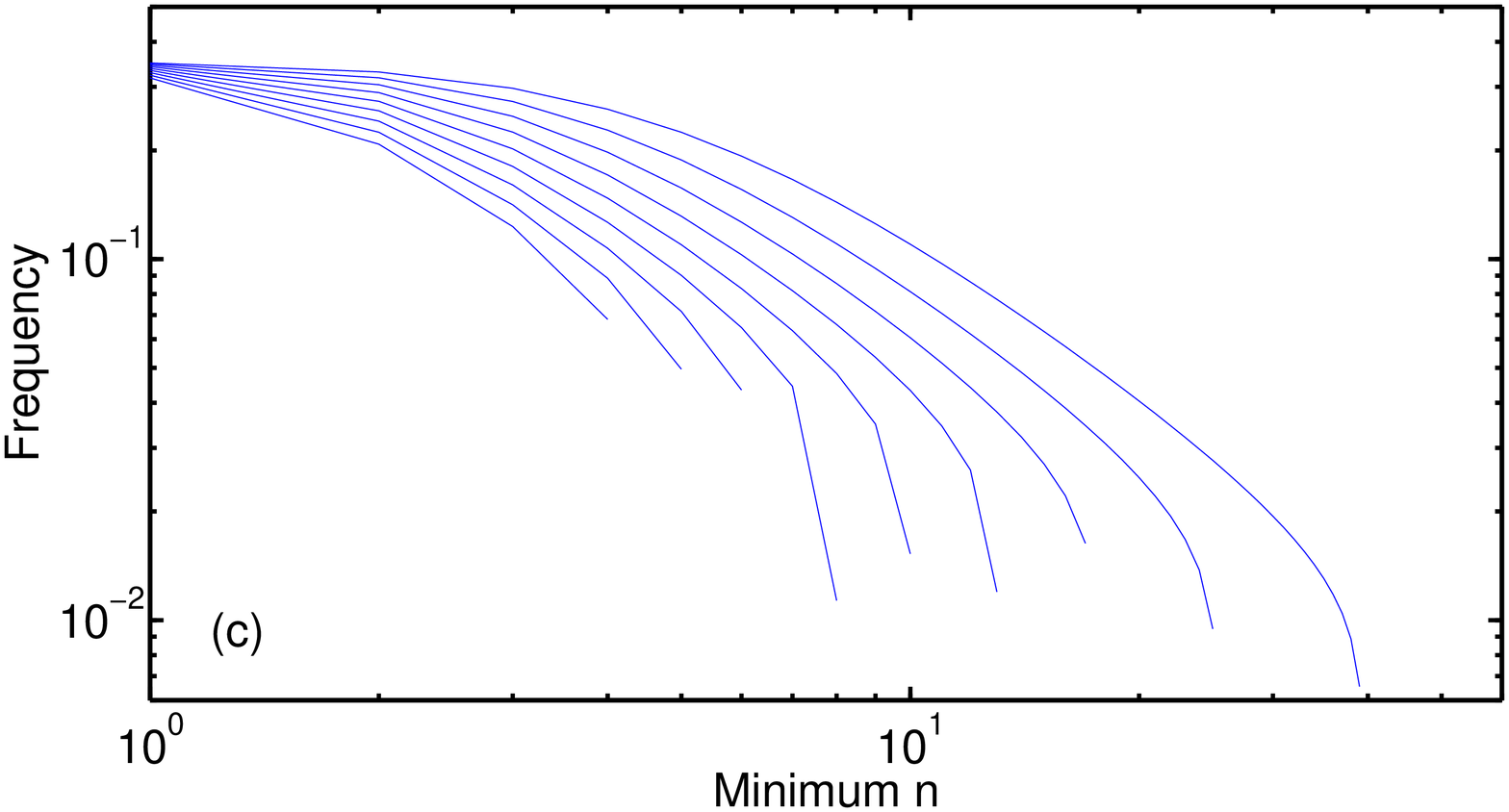}
\caption{\label{fig:fig2} Effective potential and phase space. (a) $V_{eff}$ as a function of $\Phi = \frac{s}{2 \pi \alpha}$
for $R=1$ and $g=1$. Bottom to top curves with decreasing value of $h$ show an increasing number of stable equilibria and 
corresponding potential wells.
The five lowest curves (from bottom) correspond to $h=5.0, 2.92, 1.89, 1.51, 1.29$ and the top two curves belong to $h= 0.48, 0.20$.
The inset shows the position $\frac{s}{n\pi \alpha}$ of the alternating sequence of minima and maxima
as a function of the enumerated minima and maxima. (b) Phase space $(s,p_s=\frac{m}{2} {\dot{s}})$ cutout ($m=2, h=1.18$). The inset shows the number of minima
of $V_{eff}$ as a function of $h$ in a double logarithmic representation. (c) Harmonic frequencies of the potential wells depending
on the minimum $n$ (counted from the minimum closest to the origin). Curves from top to bottom belong to $h=0.4$ up to $h=1.2$ in steps of $0.1$,
respectively.}
\end{figure}

In particular, the above equations require $h \le 2 \pi R$, i.e. the pitch has to be smaller than the circumference
of the projected circle in order to obtain extrema in the potential energy. As a result of the above conditions
the number of extrema and in particular minima of the potential $V_{\mathrm{eff}}$ can be tuned by adjusting the
parameters $R,h$. Figure 2(a) shows the effective potential for different values of $h$. An increasing
number of stable equilibrium configurations and consequently potential wells which lead to bound states
are encountered with decreasing value of $h$.
This number is in principal unbounded for $h \rightarrow 0$. The depth of these wells increases with decreasing
$h$ and their position corresponding to the relative position of the two particles decreases slightly. Decreasing $h$ means that we have closer near encounters
of the helix (see Fig.1), when moving onward from a given position of the helix. It is exactly this spatial recurrence
of the helix which leads to the occurence of the potential wells and which turns an overall repulsive interaction
in three dimensions into an oscillatory long-range interaction on the helix. As a consequence two repulsively
interacting particles in 3D possess in general a certain number of stable equilibria around which a bounded oscillatory
motion takes place. The two particles can therefore travel in (different) bound states with a free center of mass
motion through the helix. This is further illustrated in Fig 2(b) which shows the phase portrait $(s,p_s)$ belonging to the Lagrangian (\ref{twobodham}).
Clearly visible are the stable fixed points, the bounded librational motion around them and the scattering winding trajectories
coming from and going to infinity. Note that in case of attractive interactions the sign of $V_{eff}$
has to be reversed and consequently the minimum/maximum configurations have to be exchanged when compared to the case $g > 0$. 
The inset in Figure 2(a) shows how the stable fixed points, i.e. the minima, are
distributed over the helix. The relative phases $\phi=\frac{s}{\alpha n \pi}$ of the positions of the two particles 
closely fluctuate in a regular manner around $1$ with a slight increase of the fluctuations for increasing distance
of the particles. This indicates that the minima and maxima of the potential energy are distributed on the opposite
and same sides of the helical winding (see Fig. 1). The step function corresponding to the number of minima of $V_{\mathrm{eff}}$
as a function of $h$ is shown in the inset of Fig. 2(b). It exhibits a power law behaviour with a (negative) quadratic exponent  
which can be inferred from eq.(\ref{cond1}) in the limit $h \rightarrow 0$. The emergence of new minima and maxima
with decreasing $h$ happens via a saddle node bifurcation cascade which, of course, obeys the same scaling behaviour 
as the number of minima.

Figure 2(c) shows the dependence of the frequencies of the oscillatory dynamics or bound states around the stable equilibria on
the order of the minimum (counting from the one closest to the origin) within an harmonic approximation
for different values of the pitch $h$. The frequencies monotonically
decrease for minima with increasing distance between the particles, i.e. the potential wells become increasingly shallower
covering two orders of magnitude with respect to their frequencies in the given example.
As one would expect, the frequencies are found to be overall larger for smaller values of $h$.
The ratios of the frequencies belonging to different values of $h$ increase with increasing order $n$ of the
minimum. From the above we can conclude that the effective interaction potential is to a large extent tunable, i.e.
the number of minima, their frequencies and positions can be adjusted by varying the underlying parameters $R,h$.

\begin{figure}
\includegraphics[width=0.5\columnwidth]{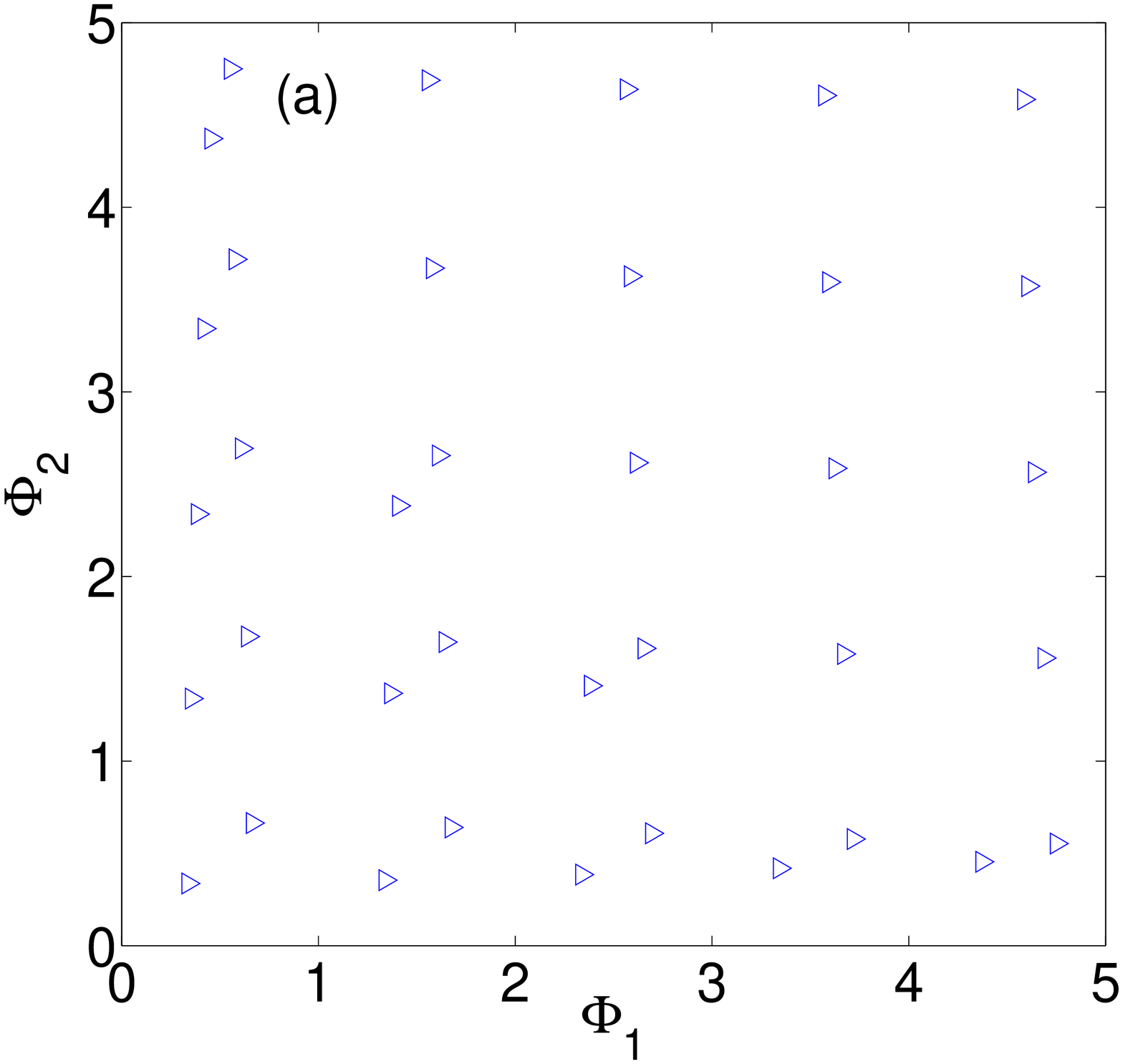} \hspace*{-0.7cm}
\includegraphics[width=0.5\columnwidth]{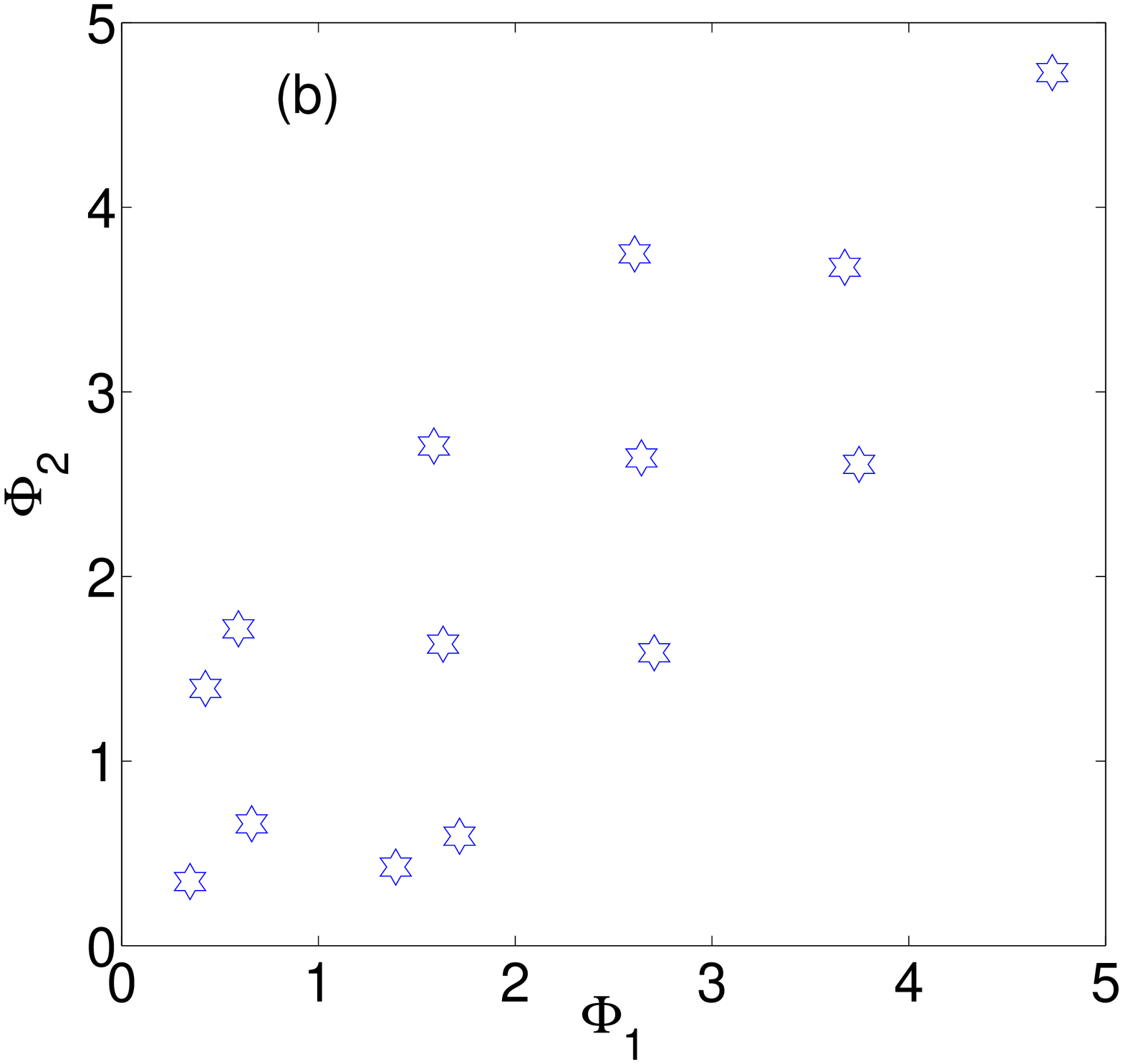}\\
\includegraphics[width=0.5\columnwidth]{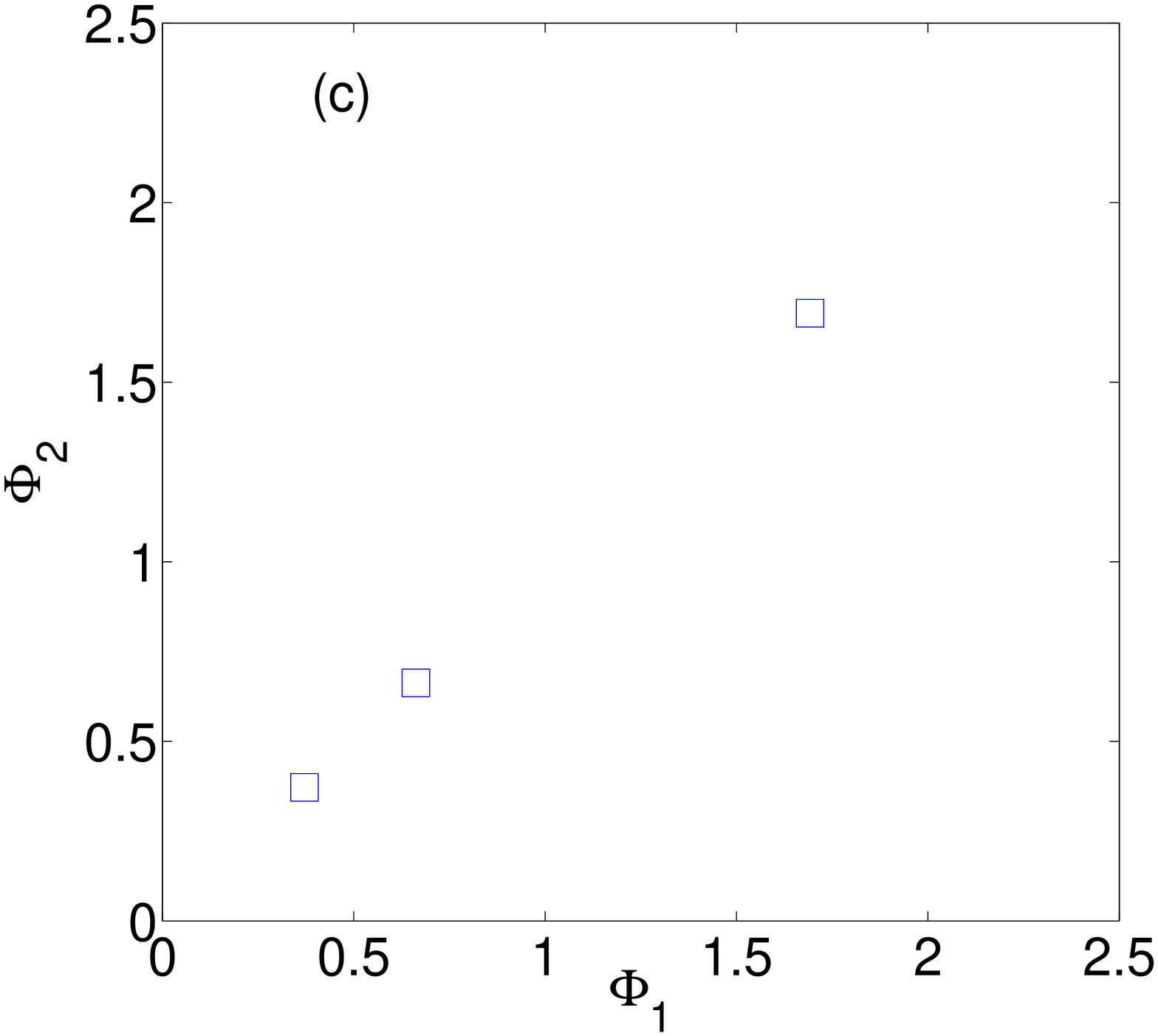} \hspace*{-0.7cm}
\includegraphics[width=0.5\columnwidth]{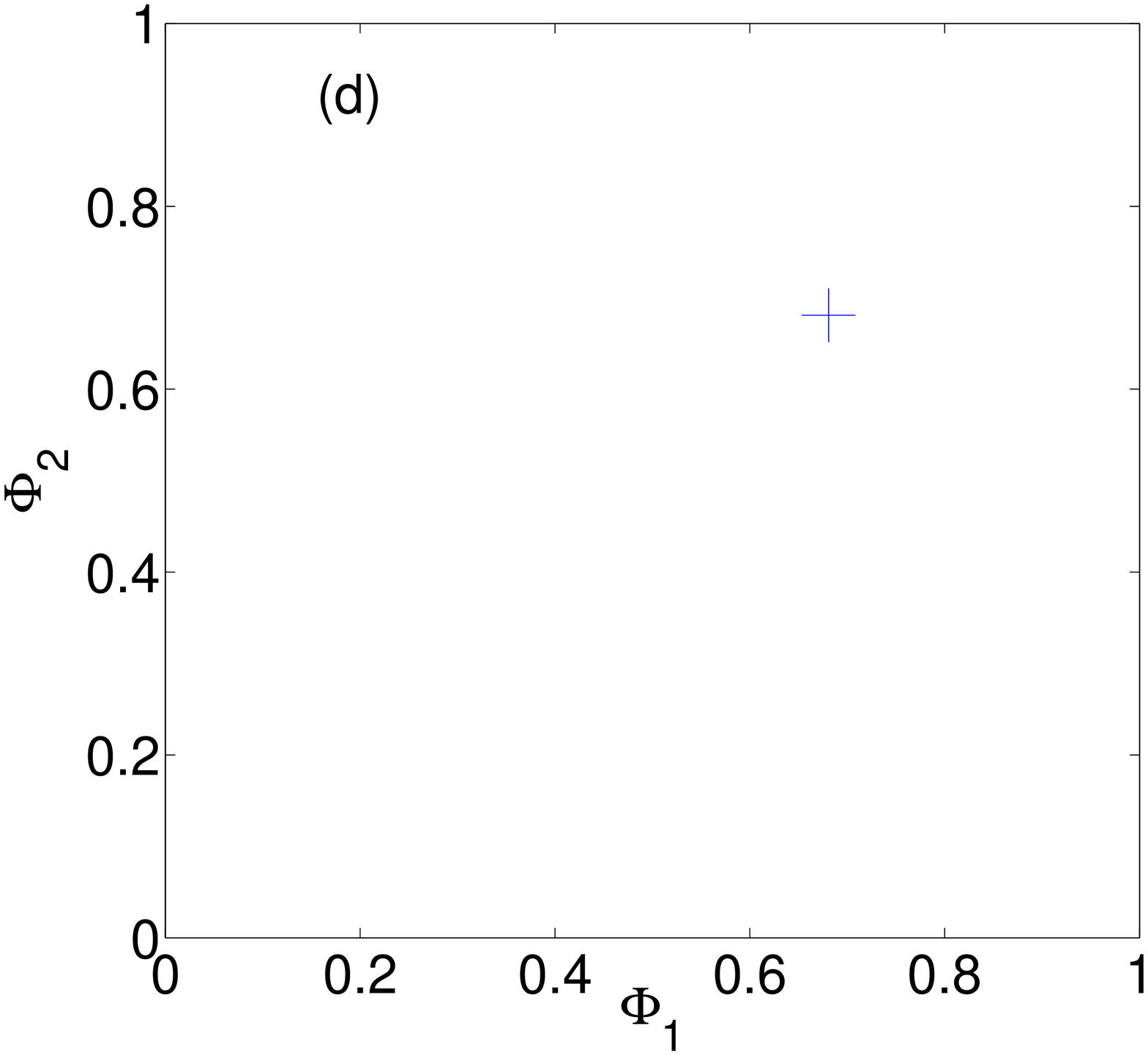}\\
\includegraphics[width=0.85\columnwidth]{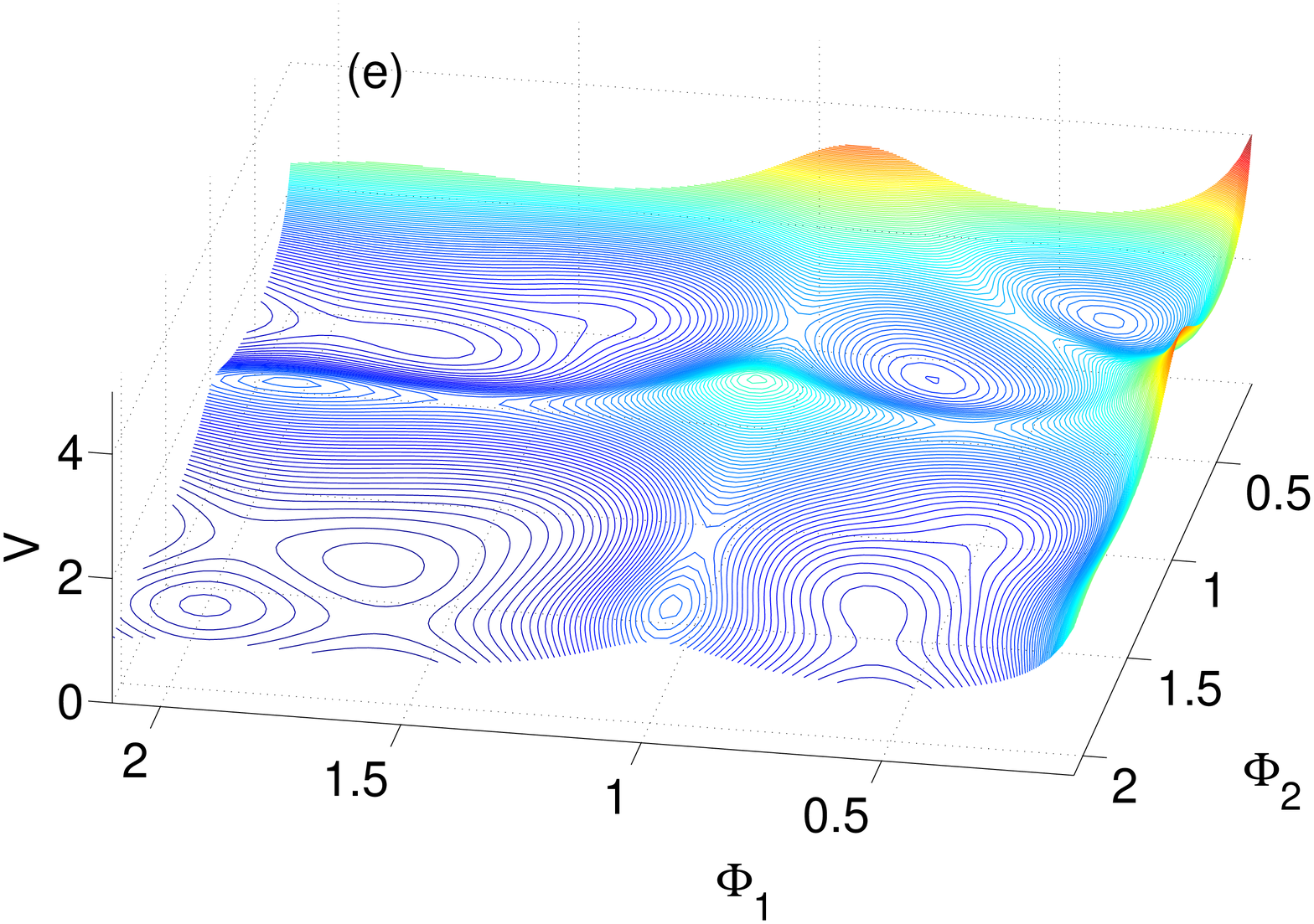}
\caption{\label{fig:fig5} (a-d) Distribution of equilibrium configurations in $(\Phi_1,\Phi_2)$-space ($\Phi_i=\frac{s_i}{\alpha 2 \pi}$)
yielding confining potential wells and bound states for three-particles in the helix. Values for the pitch $h$ are
$0.5$ (triangles), $1.0$ (stars), $1.5$ (squares) and $2.0$ (crosses), respectively for the different subfigures. (e) Three-dimensional contour plot
of the potential energy for three particles for $h=1.1$. }
\end{figure}

\begin{table*}
\begin{minipage}{9.6cm}
\begin{tabular}{|l|l|l|}\hline\hline
        &  $ N = 4$                             & $N=5$               \\ 
        &  \hspace*{1mm} $\Phi_1$ \hspace*{3mm} - $\Phi_2$ \hspace*{3mm} - $\Phi_3$  & \hspace*{1mm} $\Phi_1$ \hspace*{3mm} - $\Phi_2$ \hspace*{3mm} - $\Phi_3$ \hspace*{3mm} - $\Phi_4$                       \\ \hline
$h=1.1$  & 0.273 -  0.267 -  0.273                 & 0.261 - 0.357 - 0.256 - 0.557 \\ 
        & 0.303 - 0.473 - 0.750                   & 0.274 - 0.574 - 0.574 - 0.273 \\
        & 0.471 - 0.292 - 0.471                   &  0.361 - 0.239 - 0.239 - 0.361                      \\
        & 0.750 - 0.473 - 0.303                   &  0.328 - 0.458 - 0.790 - 0.637                      \\
        & 0.681 - 0.733 - 0.681                 &  0.418 - 0.392 - 0.392 - 0.418                      \\
        & 0.667 - 0.689 - 1.766                 & 0.483 - 0.268 - 0.553 - 0.737                      \\
        & 1.766 - 0.689 - 0.667                 & 0.557 - 0.256 - 0.357 - 0.261                      \\
        & 1.754 - 0.632 - 1.754                 & 0.669 - 0.754 - 0.754 - 0.669                       \\
        & 1.635 - 1.727 - 1.635                 & 0.737 - 0.553 - 0.268 - 0.483                      \\ 
        &                                       &  0.780 - 0.416 - 0.417 - 0.780                                      \\ 
        &                                       &  0.637 - 0.790 - 0.458 - 0.328                                     \\ 
        &                                       &  0.307 - 0.477 - 0.768 - 1.757                                     \\ 
        &                                       &  0.681 - 0.743 - 0.715 - 1.756                                      \\ 
        &                                       & 1.756 - 0.715 - 0.743 - 0.681                                       \\ 
        &                                       & 1.644 - 1.739 - 1.739 - 1.644                                      \\ 
        &                                       & 1.790 - 0.699 - 0.699 - 1.790                                      \\ 
        &                                       &  1.757 - 0.768 - 0.477 - 0.307                                     \\  \hline
$h=1.5$ & 0.470 - 0.327  - 0.470   & 0.670 - 0.768 - 0.768 - 0.670  \\
        & 0.672 - 0.748 -  0.672                 &  \\ 
        &                  &  \\ 
        &                  &  \\  \hline
\end{tabular}
\caption{Four and five-particle equilibrium configurations corresponding to bound states on the helix for two different values of the 
pitch. Positions on the helix are provided in terms of $\Phi_i = \frac{s_i}{2 \Pi \alpha}$}
\end{minipage}
\end{table*}                                                                                                                                                     

\section{Many particle systems}

Proceeding to more particle systems, the potential energy landscape $V_{\mathrm{eff}}$
rapidly develops a large number of local minima with decreasing value of the pitch (see Fig.3(e)). The two limiting cases $h \rightarrow 0$ and
$h \rightarrow \infty$ are evident. The first case corresponds to placing the particles in an equidistant manner on a ring and
the second one to a zero curvature straight line and accordingly no confining wells occur, i.e. we encounter the 1D Coulomb interaction.
In between these limiting cases the interesting physics happens. Figure 3(a-d) shows the distribution of local minima in the $\Phi_1,\Phi_2$-plane
with varying value of the pitch where $\Phi_i=\frac{s_i}{2 \pi \alpha}, i=1,2$ represent the two normalized distances for the three particles.
For $h=2.0$ (Fig.3(d)) a single equilibrium configuration of equidistantly placed particles with $\Phi_{1,2} \approx 0.68$ 
exists. The latter means that the three particles cover $1.5$ windings of the helix. For $h=1.5$ (Fig.3(c))
already three stable equilibrium configurations and
correspondingly bound states exist all of them being equidistant $\Phi_1 = \Phi_2$ and covering one, two  and a little more than three windings,
respectively. For $h=1.0$ (Fig.3(b)) and $h=0.5$ (Fig.3(a)) we encounter $14$ respectively $37$ equilibria in the considered
coordinate regime, now many of them being of asymmetric character (see in particular
Fig.3(a)). Fig. 3(e) illustrates the complexity of the potential energy landscape already manifest for a comparatively high value $h=1.1$.
Finally, table I shows all four and five particle equilibria for $h=1.5$ and $h=1.1$. For the higher value $h=1.5$ only configurations symmetric
with respect to the midpoint of the four and five-particle chain occur. For $h=1.1$ however, symmetric and asymmetric configurations are found, the
latter being of increasing or decreasing order or even oscillating with respect to the values of $\Phi_i, i=1,2,3$.

Extrapolating from a few to many-particle systems the potential landscape of ions in a helical chain will exhibit a plethora of local minima
reminescent of spin glasses \cite{Binder} which are characterized by frustrated interactions and stochastic disorder. The metastable structures in spin glasses
are replaced here by the metastable ionic pair configurations where two ions are located at different windings of the helix and separated by
a finite potential barrier. With increasing distance (in terms of windings) between the particles this barrier lowers and the many-particle
quantum helical chain therefore comprises many metastable structures with largely varying time-scales. These facts will presumably lead to
intriguing thermodynamical properties of our helical chain of ions. The non-ergodic dynamics of spin glasses for low temperature occurs
due to trapping in the deep tales of the hierarchically frustrated energy landscape. In our case of the ionic helical chain we also expect
non ergodic behaviour. However, in contrast to spin glasses the minima and corresponding metastable states follow a hierarchical scheme.
Indeed, Monte-Carlo simulations of the statistical finite-temperature
properties of the classical chain indicate already a very unusual low-temperature behaviour below and at the transition point where the
ions can pass the corresponding potential barriers i.e. traverse the blocked windings of the chain. However, a detailed study of these properties goes beyond the scope of the
present work and will be presented elsewhere \cite{Zampetaki}.

\section{Conclusions}

By using the example of the Coulomb interaction in three dimensions we have shown that a confinement of the motion to a one-dimensional curved
manifold introduces a novel type of effective interaction among the particles. The latter possesses a peculiar form unknown from fundamental
interactions in Cartesian space and is to a large extent tunable by changing the curvature or pitch of the underlying helical configuration. 
Due to the helical winding, the repulsive Coulomb interaction can develop for sufficiently small pitch an arbitrary number of local equilibrium
configurations and corresponding bound states which have been explored in detail for few-body systems.
It is evident that for many-particle chains there exist a plethora of local equilibria covering a wide regime of interparticle
separations. It is then a natural step to develop the quantum physics of the many-body interacting helical chain.
This will lead to novel structural properties such as enriched phase diagrams as well as an intriguing dynamical behaviour
of such a chain. Open questions include the metastability of the observed equilibria and the quest for the existence of long-range
versus short-range order in these systems. The analogy to spin glasses with their frustrated interactions and many metastable states
on varying time scales suggests an equally intricate statistical finite temperature behaviour to occur for our many-body helical chain.

Finally, let us briefly address the question of possible experimental realizations of the helix topology. Ion trap technology and in
particular microfabrication of ion traps \cite{hughes} offer a wide range of possibilities to control even individual ions and manipulate
them with electric fields as well as optical and microwave radiation. Still, the possibility to create a helical configuration in this framework represents
an experimental challenge. It can however be realized in the context of free-standing nano-objects based on few monolayers thick
scrolled heterostructures \cite{prinz} leading to helices where both the radius and pitch are of the order of 10-100 nm.
For neutral species, on the other hand side, optical interference based chiral microstructures can be fabricated using
holographic lithography \cite{pang}. Also the work in ref.\cite{law} clearly demonstrates the experimental relevance and realizability
as well as the novel physics of dipolarly interacting molecules in helical optical lattices.
An intriguing perspective is here the use of evanescent fields surrounding optical nanofibers which allow to trap and coherently
process neutral atoms or polar molecules \cite{vetsch}.
\vspace*{0.3cm}

\acknowledgments
The author thanks A. Rauschenbeutel and F. Lenz for a careful reading of the manuscript and valuable comments.
Discussions with F.K. Diakonos are gratefully appreciated.

\end{document}